\documentclass[preprintnumbers,amsmath,11pt,amssymb,floatfix,superscriptaddress,showpacs,nofootinbib]{revtex4}
\usepackage{graphicx}
\usepackage{epsfig}
\usepackage{bm}
\usepackage{amsfonts}

\input epsf

\begin{document}

\title{\Large Behaviour of interacting Ricci dark energy in \emph{logarithmic} $f(T)$ gravity}

\author{Rahul Ghosh}
\email{ghoshrahul3@gmail.com} \affiliation{Department of
Mathematics, Bhairab Ganguly College, Kolkata-700 056, India.}

\author{Surajit Chattopadhyay}
\email{surajit_2008@yahoo.co.in, surajcha@iucaa.ernet.in}
\affiliation{ Pailan College of Management and Technology, Bengal
Pailan Park, Kolkata-700 104, India.}

\date{\today}

\begin{abstract}
In the present work we have considered a modified gravity dubbed
as ``logarithmic $f(T)$ gravity" \cite{Bambaetal} and investigated
the behavior of Ricci dark energy interacting with pressureless
dark matter. We have chosen the interaction term in the form
$Q\propto H\delta\rho_{m}$ and investigated the behavior of the
Hubble parameter $H$ as a function of the redshift $z$. For this
reconstructed $H$ we have investigated the behavior of the density
of the Ricci dark energy $\rho_{RDE}$ and density contribution due
to torsion $\rho_{T}$. All of the said cosmological parameters are
seen to have increasing behavior from higher to lower redshifts
for all values of $c^{2}$ pertaining to the Ricci dark energy.
Subsequently, we observed the equation of state parameter
$w_{RDE}$ in this situation. The equation of state parameter is
found to behave like phantom for all choices of $c^{2}$ in the
Ricci dark energy.
\end{abstract}


\maketitle

\section{Introduction}

Accelerated expansion of the universe is well-established by the
works of \cite{Spergel, Perlmu}. The ``dark energy" (DE),
characterized by negative pressure, is responsible for this cosmic
acceleration \cite{Copeland, Paddy, BambaOdintsov, Tsuza}.
Importance of modified gravity for late acceleration of the
universe has been reviewed by \cite{Nojiri1, Clifton}.  Various
modified gravity theories have been proposed so far. These
include, $f(R)$ \cite{Nojiri2, Nojiri3}, $f(T)$ \cite{Cai1,
Ferraro, Bamba1,BambaGeng}, $f(G)$ \cite{Myrza2, Setare1},
Horava-Lifshitz \cite{Kiritsis, Nishioka} and Gauss-Bonnet
\cite{Nojiri4, Li} theories. One of the newest extended theories
of gravity is the so-called $f(T)$ gravity, which is a theory
formulated in a spacetime possessing absolute parallelism
\cite{Ferraro}. Some fundamental aspects of $f(T)$ theories have
been studied in the works of \cite{Li1} and \cite{Sotioru}. In
this theory of modified gravity, the teleparallel Lagrangian
density described by the torsion scalar $T$ has been promoted to a
function of $T$, i.e., $f(T)$, in order to account for the late
time cosmic acceleration \cite{bambageng, myrza3}.

Models of  DE include quintessence \cite{Ratra}, phantom
\cite{Najiri}, quintom \cite{saridakis}, Chaplygin gas
\cite{Gorini}, tachyon \cite{chimento}, hessence \cite{Zhao} etc.
All DE models can be classified by the behaviors of equations of
state as following \cite{saridakis}:(i)Cosmological constant: its
EoS is exactly equal to $-1$, that is $w_{DE}=-1$;
(ii)Quintessence: its EoS remains above the cosmological constant
boundary, that is $w_{DE}\geq-1$; (iii)Phantom: its EoS lies below
the cosmological constant boundary, that is $w_{DE}\leq-1$ and
(iv)Quintom: its EoS is able to evolve across the cosmological
constant boundary. Inspired by the holographic principle
\cite{Hsu, Xhang}, Gao et al.\cite{Gao} took the Ricci scalar as
the IR cut-off and named it the Ricci dark energy (RDE), in which
they take the Ricci scalar $R$ as the IR cutoff. With proper
choice of parameters the equation of state crosses $-1$, so it is
a `quintom' \cite{Feng}. The Ricci scalar of FRW universe is given
by $R=-6\left(\dot{H} +2H^{2}+\frac{k}{a^{2}}\right)$, where $H$
is the Hubble parameter, $a$ is the scale factor and $k$ is the
curvature. The energy density of RDE is given by
$\rho_{RDE}=3c^{2}\left(\dot{H}+2H^{2}+\frac{k}{a^{2}}\right)$. In
flat FRW universe, $k=0$ and hence
$\rho_{RDE}=3c^{2}\left(\dot{H}+2H^{2}\right)$.

Interacting DE models have gained immense interest in recent
times. Works in this direction include \cite{setare1, setare2,
setare3, Suwa, jamil1, jamil2}. In a recent work by Jamil et al
\cite{jamil1} examined the interacting dark energy model in $f(T)$
cosmology assuming dark energy as a perfect fluid and choosing a
specific cosmologically viable form $f(T)=\beta \sqrt{T}$.
Interacting RDE was considered in \cite{Suwa}, where the
observational constraints on interacting RDE were investigated. In
another recent work, Pasqua et al \cite{Pasqua} reconstructed the
potential and the dynamics of the tachyon, K-essence, dilaton and
quintessence scalar field models according to the evolutionary
behavior of the interacting entropy-corrected holographic RDE
model. In the present work, we have considered an interacting RDE
in the ``logarithmic $f(T)$ gravity" proposed by \cite{Bambaetal}.
In the said form of $f(T)$ gravity, the form of $f(T)$ is proposed
as $f(T)=\beta
T_{0}\left(\frac{qT_{0}}{T}\right)^{-1/2}\ln\left(\frac{qT_{0}}{T}\right)$,
where $\beta=\frac{1-\Omega_{m}^{(0)}}{2q^{-1/2}}$.

This logarithmic $f(T)$ gravity model is basically constructed
based on phenomenological approach. Therefore, the motivation to
examine this model is that if we consider the interaction between
the Ricci dark energy and dark matter in this model, we can obtain
some desirable cosmological consequence.
\\
\section{An overview of $f(T)$ gravity}
In the framework of $f(T)$ theory, the action of modified TG is
given by \cite{myrza3}
\begin{equation}
I=\frac{1}{16\pi G}\int d^{4}x \sqrt{-g}\left[f(T)+L_{m}\right]
\end{equation}
where, $L_{m}$ is the Lagrangian density of the matter inside the
universe. We consider a flat Friedmann-Robertson-Walker (FRW)
universe filled with the pressureless matter. Choosing $(8\pi
G=1)$ the modified Friedman equations in the framework of $f(T)$
gravity are given by \cite{Ferraro, myrza3}
\begin{equation}
H^{2}=\frac{1}{3}\left(\rho+\rho_{T}\right)
\end{equation}
\begin{equation}
2\dot{H}+3H^{2}=-\left(p+p_{T}\right)
\end{equation}
where,
\begin{equation}
\rho_{T}=\frac{1}{2}(2T f_{T}-f-T)
\end{equation}
\begin{equation}
p_{T}=-\frac{1}{2}\left[-8\dot{H}T
f_{TT}+(2T-4\dot{H})f_{T}-f+4\dot{H}-T\right]
\end{equation}
Where \cite{Ferraro}
\begin{equation}
T=-6\left(H^{2}\right)
\end{equation}
As we are considering interaction between pressureless dark matter
and RDE, we shall have $\rho=\rho_{m}+\rho_{RDE}$ and $p=p_{RDE}$
in the equations (2) and (3). In the lograrithmic $f(T)$ gravity
\cite{Bambaetal}
\begin{equation}
f(T)=\beta
T_{0}\left(\frac{qT_{0}}{T}\right)^{-1/2}\ln\left(\frac{qT_{0}}{T}\right)
\end{equation}
where, $\beta=\frac{1-\Omega_{m}^{(0)}}{2q^{-1/2}}$ and $q$ is a
positive constant. In \cite{Bambaetal}, Bamba et al. have shown
that for the said form of $f(T)$ gravity, the EoS parameter stays
above $-1$, when plotted against redshift $z$. In the present
work, we shall investigate the behavior of the EoS parameter when
interacting RDE is considered in the said form of modified
gravity. In the next section, we shall briefly describe the
mathematical background of the RDE.

\section{Interacting RDE}
The metric of a spatially flat homogeneous and isotropic universe
in FRW model is given by
\begin{equation}
ds^{2}=dt^{2}-a^{2}(t)\left[dr^{2}+r^{2}(d\theta^{2}+sin^{2}\theta
d\phi^{2})\right]
\end{equation}
where $a(t)$ is the scale factor. The conservation equation is
given by
\begin{equation}
\dot{\rho}+3H(\rho+p)=0
\end{equation}
As we are considering interaction between RDE and dark matter, the
conservation equation will take the following form
\begin{equation}
\dot{\rho}_{total}+3H(\rho_{total}+p_{total})=0
\end{equation}
where, $\rho_{total}=\rho_{RDE}+\rho_{m}$ and $p_{total}=p_{RDE}$
(as we are considering pressureless dark matter, $p_{m}=0$). It is
already stated that \cite{Gao}
\begin{equation}
\rho_{RDE}=3c^{2}\left(\dot{H}+2H^{2}+\frac{k}{a^{2}}\right)
\end{equation}
As in the case of interaction the components do not satisfy the
conservation equation separately, we need to reconstruct the
conservation equation by introducing an interaction term $Q$.
Considering the interaction term $Q$ as $Q=3H\delta\rho_{m}$
\cite{sheykhi, cataldo}, where $\delta$ is the interaction
parameter, the conservation equation takes the form

\begin{equation}
\dot{\rho}_{RDE}+3H(\rho_{RDE}+p_{RDE})=3H\delta\rho_{m}
\end{equation}
and
\begin{equation}
\dot{\rho}_{m}+3H\rho_{m}=-3H\delta\rho_{m}
\end{equation}
In the subsequent section, we shall discuss the behaviors of the
various cosmological parameters when the interacting RDE is
considered in the logarithmic $f(T)$ gravity.
\section{Discussion}
Solving the conservation equation (15) we get under interaction
\begin{equation}
\rho_{m}=\rho_{m0}a^{-3(1+\delta)}
\end{equation}
Using equations (6) and (7) we get $\rho_{T}$ from equation (4)
and use it in (2) along with (14) and get the following
differential equation
\begin{equation}
(15+6c^{2})H^{2}+\left(\frac{6\beta
H_{0}}{\sqrt{q}}\right)H+3c^{2}\dot{H}+\rho_{m0}a^{-3(1+\delta)}=0
\end{equation}
From equation (15) we get the new $H$ under the said interaction
considered in logarithmic $f(T)$ gravity and plot against redshift
$z=a^{-1}-1$ in figure 1 for $c^{2}<0.5,~=0.5$ and $>0.5$. The
three cases are indicated by solid, dashed and dotted lines
respectively in the figure. This is followed in the subsequent
figures also. Following \cite{Bambaetal}, we take
$\Omega_{m}^{(0)}=0.26$ while plotting $H$. As we approach from
higher to lower redshifts, we observe that $H(z)$ is exhibiting an
increasing pattern. The rate of increase is sharper in the case of
$c^{2}<0.5$ than the other two cases.
\begin{figure}
\includegraphics[height=2.8in]{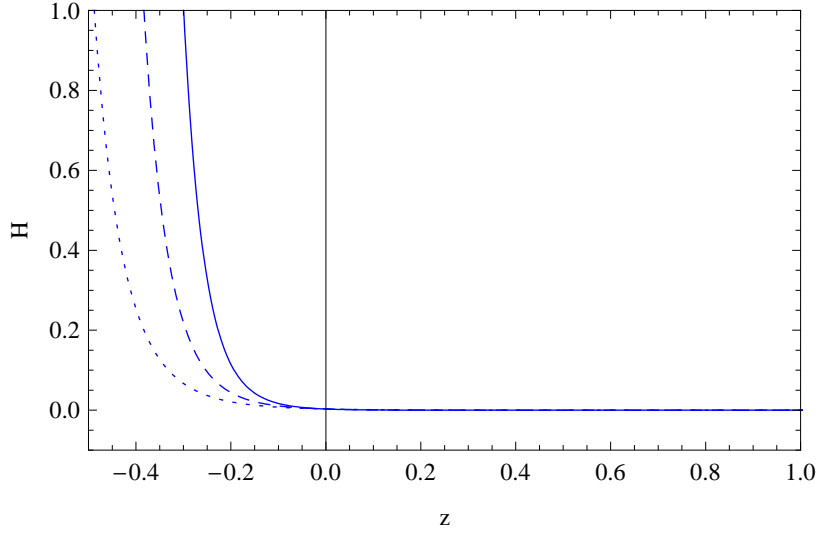}\\
\caption{ This figure plots the Hubble parameter $H$ for
interacting RDE under logarithmic $f(T)$ gravity. We have taken
$\Omega_{m}^{0}=0.26$ and $\delta=0.05$.}
\end{figure}
To view the behavior of the logarithmic $f(T)$ we have plotted it
against $z$ in figure 2.
\begin{figure}
\includegraphics[height=2.8in]{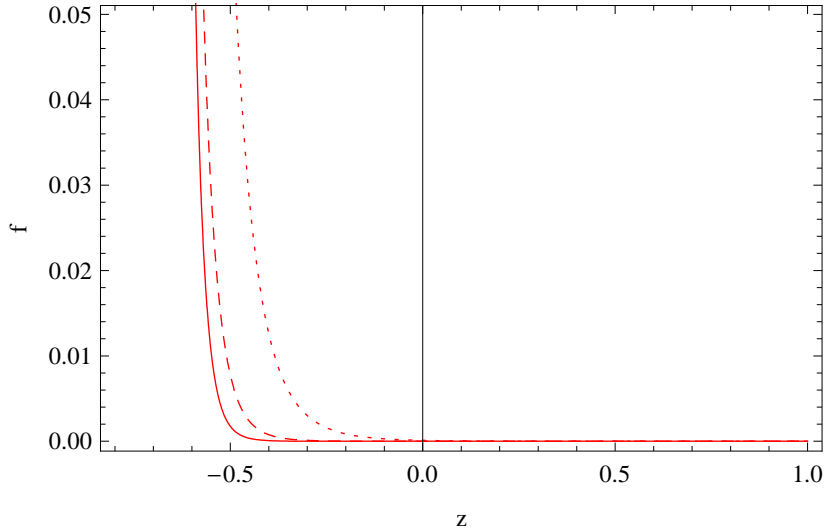}\\
\caption{This figure plots the $f(T)$ against redshift $z$. We
have taken $\Omega_{m}^{0}=0.26$ and $\delta=0.05$.}
\end{figure}
From figure 2 we observe that $f(T)$ is increasing with evolution
of the universe.
\begin{figure}
\includegraphics[height=2.8in]{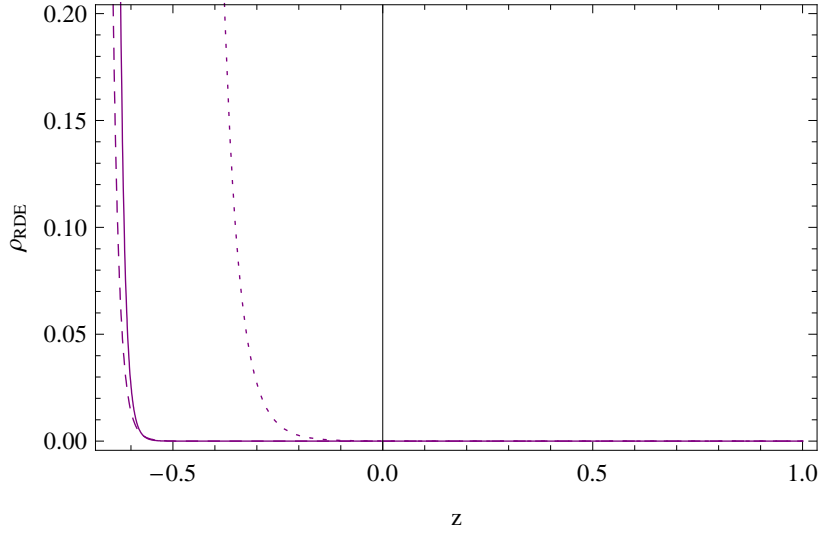}\\
\caption{This figure plots the $\rho_{RDE}$ against redshift $z$.
We have taken $\Omega_{m}^{0}=0.26$ and $\delta=0.05$.}
\end{figure}
\begin{figure}
\includegraphics[height=2.8in]{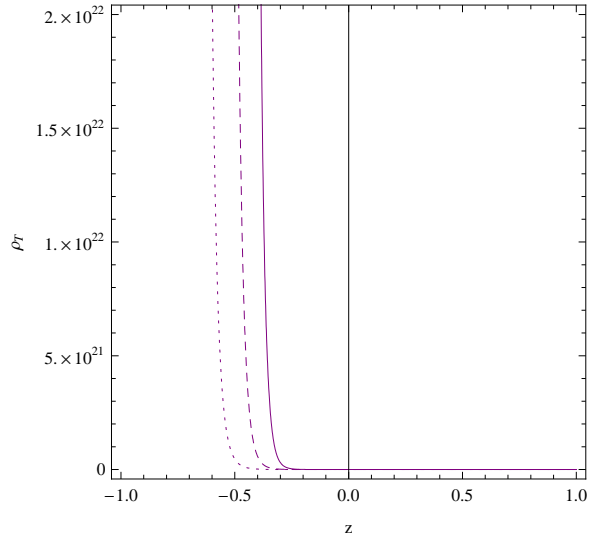}\\
\caption{This figure plots the density contribution due to torsion
i.e. $\rho_{T}$ against redshift $z$. We have taken
$\Omega_{m}^{0}=0.26$ and $\delta=0.05$.}
\end{figure}
\begin{figure}[h]
\begin{minipage}{14pc}
\includegraphics[width=16pc]{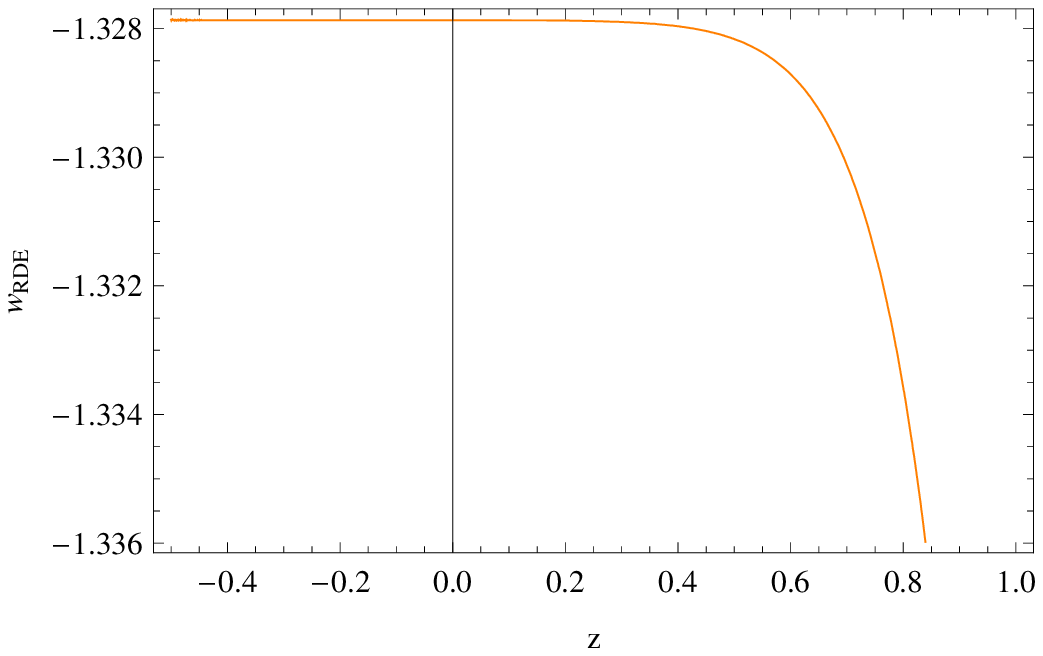}
\caption{\label{label}Evolution of $w_{RDE}$ for $c^{2}<0.5$.}
\end{minipage}\hspace{3pc}%
\begin{minipage}{14pc}
\includegraphics[width=16pc]{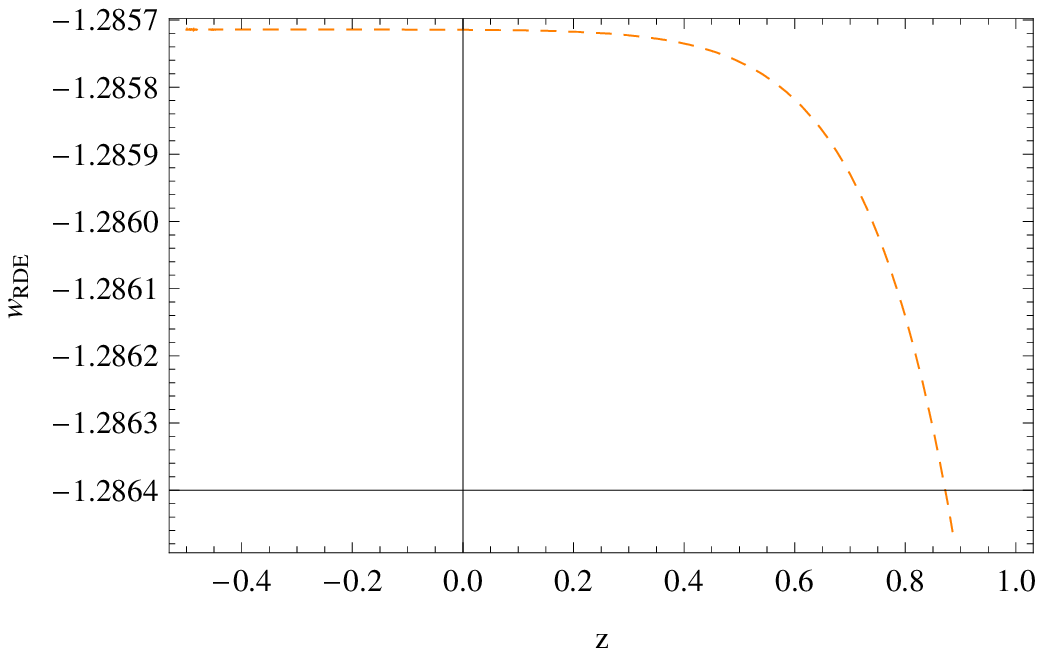}
\caption{\label{label}Evolution of $w_{RDE}$ for $c^{2}=0.5$.}
\end{minipage}\hspace{3pc}%
\begin{minipage}{14pc}
\includegraphics[width=16pc]{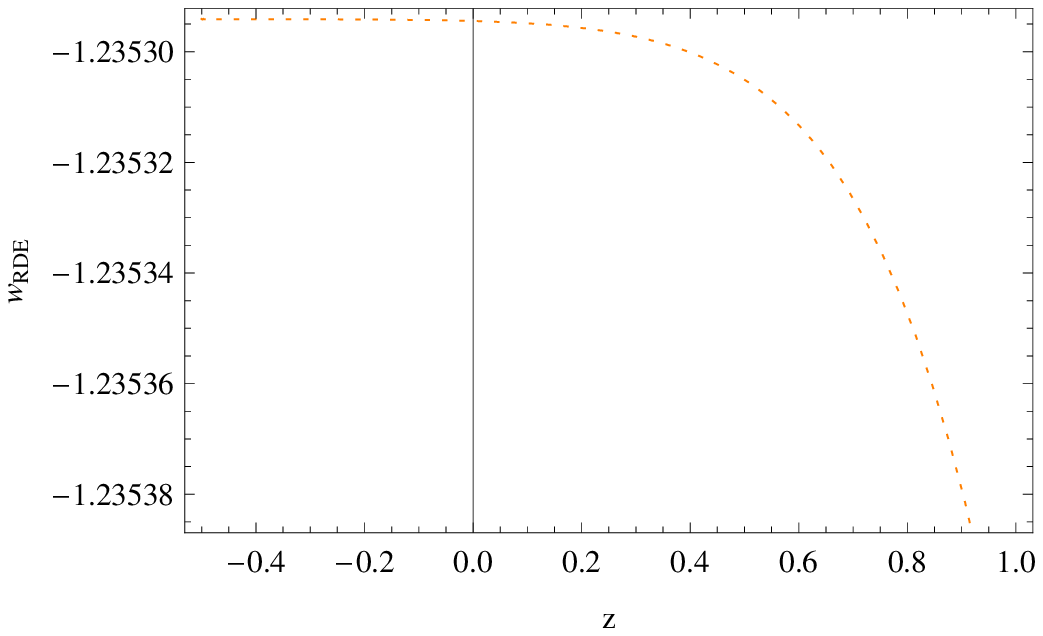}
\caption{\label{label}Evolution of $w_{RDE}$ for $c^{2}>0.5$.}
\end{minipage}\hspace{3pc}%
\end{figure}

In figure 3, we have plotted the density $\rho_{RDE}$ under the
interaction in the proposed model of $f(T)$ gravity. We observe
that for all of the choices of $c^{2}$, the density of RDE is
increasing with the evolution of the universe. This is consistent
with the fact that the universe has evolved from matter dominated
to DE dominated phase. In figure 4, we plot the density
contribution due to torsion and we observe that like $\rho_{RDE}$,
the $\rho_{T}$ is increasing as the universe evolves from higher
to lower redshifts. In figures $5-7$ we have plotted the EoS
parameter for $c^{2}<0.5,~=0.5$ and $>0.5$. In all of the figures
we observe $w_{RDE}<-1$ that indicates phantom-like behavior
\cite{Cai}.
\\
\section{Concluding remarks}
In the present work we have considered a modified gravity dubbed
as ``logarithmic $f(T)$ gravity" and investigated the behavior of
Ricci dark energy interacting with pressureless dark matter. We
have chosen the interaction term in the form $Q\propto
H\delta\rho_{m}$ and investigated the behavior of the Hubble
parameter $H$ as a function of the redshift $z$. For this
reconstructed $H$ we have investigated the behavior of the density
of the Ricci dark energy $\rho_{RDE}$ and density contribution due
to torsion $\rho_{T}$. All of the said cosmological parameters are
seen to have increasing behavior from higher to lower redshifts
for all values of $c^{2}$ pertaining to the Ricci dark energy.
Subsequently, we observed the equation of state parameter
$w_{RDE}$ in this situation. In \cite{Bambaetal}, the motivation
behind this study, the logarithmic $f(T)$ gravity was found not to
cross the phantom-divide. In the present paper we considered
interacting Ricci dark energy in the said form of gravity and here
also we found that the phantom-divide can not be realized. Rather,
the equation of state parameter $w_{RDE}$ is found to stay below
$-1$ that indicates phantom-like behavior. In \cite{zhang} it was
reported that the Ricci dark energy behaves like phantom in
Einstein gravity for $c^{2}<0.5$. However, in the present work,
where we consider interacting Ricci dark energy in a modified
gravity in the form of logarithmic $f(T)$ gravity, the equation of
state parameter behaves like phantom for all choices of $c^{2}$.
\\\\
\subsection{Acknowledgement}
The second author (SC) acknowledges the research grant under Fast
Track Programme for Young Scientists provided by the Department of
Science and Technology (DST), Govt of India. The project number is
SR/FTP/PS-167/2011. Also, the second author sincerely acknowledges
Inter-University Centre for Astronomy and Astrophysics (IUCAA),
Pune, India for providing Visiting Associateship for the period of
2011-2014.
\\

\end{document}